# THEORY OF FERROMAGNETISM IN (A, Mn) B SEMICONDUCTORS

A. K. Rajagopal[a], Mogus Mochena[b], and P. J. Lin-Chung[a]
(a) Naval Research Laboratory, Washington D. C. 20375-5320
(b) Department of Physics, Florida A & M University, Tallahassee, Florida 32307

## ABSTRACT

A brief review of theory of ferromagnetism of diluted magnetic semiconductors of the form (A, Mn) B based on the double exchange model is first given. A systematic investigation of the phenomena extending the current theory is outlined. We begin with an investigation of regions of instability of the nonmagnetic towards the ferromagnetic state of a system of Mn-atoms doped in AB-type semiconductor. A self-consistent many-body theory of the ferromagnetic state is then developed, going beyond the mean field approaches by including fluctuations of the Mn-spins and the itinerant hole-gas. A functional theory suitable for computation of system properties such as Curie temperature as a function of hole and the Mn-concentration, spin-current, etc. is formulated.

## I. INTRODUCTION

Two separate technologies have been developed to date that utilize (1) semiconductor materials involving electron charge, and (2) ferromagnetic materials involving electron spin. Information technology proposals involving *both* the spin and charge of the electron in magnetic semiconductors have appeared only recently, leading to the possibility of a spin-based "*spintronics*" (Prinz 1995, 1998) technology. One of the major reasons for this delay in the practical realization and development of this may lie in the lack of control over the spin orientation as the electron is transported in any device structure, due to spin-disorder processes leading to very fast spin decoherence. Another requirement of the practicality of "*spintronics*" is that the ferromagnetism must be at room temperature. With the advent of new techniques of material fabrication and new understanding of their properties, it is hoped that the possibility to forge new technology involving spintronics may indeed be realized in the coming years.

The discovery of ferromagnetism in (In,Mn)As (Ohno et al. 1992) and (Ga,Mn)As (Ohno et al. 1996) which are ferromagnetic semiconductor alloys and many others (Dietl et al. 2000) seems to be promising in the realization of emergent tandem ferromagnet/semiconductor behaviors, in which the combination offers more possibilities than the separate features. The typical concentration of Mn is less than 10%, limited by its solubility in the host semiconductor. Experimentally, Curie temperatures are found to depend on both the hole-concentration of the host semiconductor and the Mn-content in the alloy. Based on mean field theory (Dietl et al. 2000) computed predicted values of the Curie temperatures for various p-type semiconductors containing 5% Mn and about $10^{19}$ to $10^{20}$ holes per $cm^3$, range from 20K to about 300K. These are yet to be experimentally realized. One of the most exciting experimental result is the recent work of Ohno et al (2000), who were able to control the number of holes by means of field effect transistor (FET) arrangement with (In,Mn)As. Thus electric-field control of magnetism was demonstrated which parallels the control of electric current in field effect semiconductors as in a MOSFET.

A very extensive review of the physical properties of diluted magnetic semiconductors of the type $(A^{II}, Mn) B^{VI}$ was published by Furdyna (1988). There is no similarly extensive review of the physical properties of $(A^{III}, Mn) B^{V}$ system. Mn behaves differently in these two systems because it acts both as a magnetic element and as a p type dopant. All of these are ferromagnetic below room temperature and have a band gap in the infrared and therefore are not transparent. Matsumoto et al (2001) have recently reported a room temperature ferromagnetic semiconductor based on a form of titanium dioxide, doped with a few percent of cobalt, and it has a wide enough band gap to be completely transparent. This discovery is very important in expecting a creation of "electronic paper", where in a single flat panel display, one has integrated electronic circuits and magnetic storage.

On the theoretical front, there have been several attempts in understanding the ferromagnetism of these systems. It is believed that the basic mechanism of magnetism in these materials is the interplay of the magnetic moment of the Mn atom and the itinerant hole-system of the semiconductor. This dual control of the magnetism indicates that an appropriate many-body theory of this interacting system must be constructed. The theoretical model commonly used for carrier-induced ferromagnetism in (A, Mn) B systems so far has been based on a mean field scheme of treating the "Double Exchange" process wherein the local spin of the impurity interacts with the electrons/holes of the semiconductor. This theory has not been successful in explaining the observations (Dietl et al. 2000). More investigation is called for to understand the underlying mechanism. The current state of the theory has been reviewed by Koenig et al.(2000, 2001). They also studied the spin wave excitations in these systems.

The basic model used in almost all the theoretical discussions is described by the Hamiltonian of the semiconductor hole-gas treated in effective mass approximation, (sometimes using Hubbard model) doped with Mn-atoms. The hole gas is assumed to be of low enough density so that we will at first treat them as if they are non-interacting. The effective Hamiltonian is (we use units with the Planck constant set equal to unity, $h=1$)

$$H = H_0 + H_{imp} + H_{ext}, \tag{1}$$

where

$$H_0 = \int d\vec{r} \sum_\sigma \Psi_\sigma^+(\vec{r}) \left[ -\frac{\vec{\nabla}^2}{2m^*} - \mu \right] \Psi_\sigma(\vec{r}), \tag{2}$$

$$H_{imp} = \sum_l \int d\vec{r} J(\vec{r} - \vec{R}_l) \vec{S}(\vec{R}_l) \cdot \vec{s}(\vec{r}), \tag{3}$$

Here $\vec{S}(\vec{R}_l)$ is the S=5/2 spin operator representing the Mn atom located at some position $\vec{R}_l$, and the itinerant hole-spin density is expressed in terms of the hole field operators, $\vec{s}(\vec{r}) = \frac{1}{2} \sum_{\sigma,\sigma'} \Psi_\sigma^+(\vec{r}) \vec{\tau}_{\sigma\sigma'} \Psi_{\sigma'}(\vec{r})$, where $\vec{\tau}$ is the vector of Pauli spin matrices. Here m* is the effective mass of the holes and $\mu$ is the chemical potential of the holes. We first describe the results obtained in the mean field theory, to set the stage for discussing its extension.

We employ the spinor hole-Green function to examine the itinerant magnetic property introduced by the impurity interaction, eq.(3). Thus,

$$G(1,1') = -i\langle T(\Psi_\sigma(1)\Psi^+_{\sigma'}(1'))\rangle$$

$$\equiv \begin{pmatrix} G_{\uparrow\uparrow}(1,1') & G_{\uparrow\downarrow}(1,1') \\ G_{\downarrow\uparrow}(1,1') & G_{\downarrow\downarrow}(1,1') \end{pmatrix} \qquad (4)$$

where 1 stands for space and time coordinates of the hole and $\langle L \rangle$ stands for thermal ensemble average over the system. We then write the equation of motion followed by this Green function in the standard way and express it in terms of the appropriate vertex function:

$$i\partial_{t_1} \underline{G}(1,1') = \delta(1,1')\underline{I} + \left(-\frac{\vec{\nabla}^2}{2m^*} + \mu\right)\underline{G}(1,1') + \vec{h}(1)\cdot\vec{\underline{\tau}}\,\underline{G}(1,1') +$$

$$+ \sum_{l_1} J(\vec{r}_1 - \vec{R}_{l_1})(-i)\left\langle T\left(\vec{S}(\vec{R}_{l_1},t_1)\cdot\vec{\underline{\tau}}\Psi(1)\Psi^+(1')\right)\right\rangle$$

(5)

The last term in the above equation represents the contribution of the double exchange process to the dynamics of the holes. In the mean field theory, one makes the approximation :

$$(-i)\left\langle T\left(\vec{S}(\vec{R}_{l_1},t_1)\cdot\vec{\underline{\tau}}\Psi(1)\Psi^+(1')\right)\right\rangle \cong \vec{\underline{\tau}}\cdot\left\langle \vec{S}(\vec{R}_{l_1},t_1)\right\rangle \underline{G}(1,1') \qquad (6)$$

The inverse Green function for the holes is then found to be:

$$\underline{G}^{-1}(1,1') = \underline{G}_0^{-1}(1,1') - \sum_{l_1} J(\vec{r}_1 - \vec{R}_{l_1})\left\langle \vec{S}(\vec{R}_{l_1},t_1)\right\rangle\cdot\vec{\underline{\tau}}\,\delta(1,1') \qquad (7)$$

Here $\underline{G}_0^{-1}(1.1') = \left(i\partial_{t_1} + \frac{\nabla_1^2}{2m^*} - \mu\right)\delta(1,1')\underline{I}$ is the inverse hole Green function when there is no interaction with the Mn spins, which in the Hubbard model of the semiconductor would involve the corresponding one-body Hubbard Hamiltonian. In this approximation we have retained only the effect of mean field on the holes and neglected any dynamical spin correlation function of the Mn-system. The poles of this Green function gives us the renormalized energies of the hole states. In this mean field approximation, one obtains a spin polarization of the hole system whenever the Mn-spins are themselves spin-aligned. A couple of observations on this point may be pertinent at this stage. The following phenomenology may be useful in appreciating what is being left out in such a scheme:

(a) If Mn atoms form a nonmagnetic system (paramagnetic) then we have the mean value of the Mn spins is zero while their fluctuations represented by the Mn-susceptibility is isotropic: $\langle S_z \rangle = 0$ and $\chi_{Mn\alpha,\beta} = \delta_{\alpha,\beta} S(S+1)/3$. In this case, one must go to the next order to obtain the spin polarization of the hole gas. In this second order, we have the usual RKKY (Ruderman-Kittel-Kasuya-Yoshida) type contribution to the hole self-energy.

(b) If, on the other hand, Mn atoms form a ferromagnet with $\langle S_z \rangle = S(\max)$, then since $\langle \vec{S}^2 \rangle = S(S+1)$, we have $\langle S_x^2 + S_y^2 \rangle = S(S+1) - \langle S_z^2 \rangle = S$. Then, we have the Mn-fluctuations are of the form $\chi_{Mnzz} = S^2$, and $\chi_{Mnxx} + \chi_{Mnyy} = S$. To leading order then, one just examines this expression in detail.

In such an approximation, the basic parameters of this model are then the bandwidth, W (equivalently, the density of the holes), and the exchange energy, JS, where J is the exchange integral and S is the magnitude of the spin of the atom such as Mn (S=5/2). In the indirect exchange mechanism via charge carriers, the so called RKKY interaction, is valid when W>>JS, i e., high charge density. This condition is invalid when the charge density is low where one has the opposite limit, W<<JS, where the double exchange process dominates. This type of theory results in only limited qualitative agreement with experimental results.

An exception to this type of theory is the work of Akai (1998) who performed a first principles calculation using (KKR-CPA-LDA) density functional theory to examine the carrier induced ferromagnetism in diluted magnetic semiconductor (In,Mn)As. The mechanism seems to suggest a competition between double exchange and super-exchange processes.

It appears from the descriptions of the theories of the systems under our study, one requires a theory where we need an intermediate regime. This is the open problem that needs to be resolved.

## II. OUTLINE OF OUR APPROACH

We outline here a theoretical investigation into magnetically doped semiconductors in three parts. (1) Many-body theory of the system going beyond the existing mean field theories, (2) the functional theory based on the many-body formulation which lead to a framework for numerical implementation, and finally, (3) a computation of relevant physical quantities for realistic situations, which will be predictive in character.

Before we proceed to outline of research, we present some background review of the theoretical ideas. In the 70's, Rajagopal and collaborators investigated possible ferromagnetism in 2-D systems as in Si-Mosfets, based purely on electron correlations (Rajagopal 1998). Recently there has been much activity in a somewhat similar system but now with magnetism of impurities such as Mn in semiconductors (Ohno et al. 2000). Much of the theory (Konig et al. 2000, 2001) for these systems is based on the localized spin of atoms interacting with the hole (electron) gas of a semiconductor, which leads to Mn spin-spin interaction as for example in RKKY- type mechanism. In 1964, Kochelaev developed the corresponding spin-spin interaction when the electron - gas was spin polarized, based on the theory of itinerant electron magnetism (Rajagopal 1964, 1967). In the present work, we use a semiconductor plus impurity model, and develop a theory using Green functions as in our earlier work. A CPA-type analysis of the Mn-system is then incorporated so as to arrive at a self-consistent magnetic state of this combined system.

A vector-spin-density functional theory was developed in the context of magnetism of manganites, which is superficially similar to the system under study (Rajagopal 1998). Since the functional method is a method of choice for numerical self-consistent calculations, we plan to develop an approach of this type for the doped ferromagnetic systems as well.

**A. Many-body theory**:

The system described in eqs.(1, 2, 3) consisting of interacting itinerant holes and local spins has been discussed most recently in the present context in a self-consistent mean field theory as follows. The hole system experiences the average field of the localized Mn spins, which acts as an effective magnetic field. This therefore spin-polarizes the hole gas, leading to its magnetism. On the other hand, the Mn spin in turn experiences the magnetic field due to this magnetically polarized hole system. The combined system then self-consistently goes into a ferromagnetic state. Clearly such a state depends on both the hole density and the spin-state of the Mn-system. The stability of this system is ascertained by making sure that the spin wave excitations are the first excited states of this system, which is just a collective property arising from the small, long wavelength fluctuations of the spin-polarized state around its mean polarization. This is the type of theory which ignores fluctuations of the localized Mn-spins from the start, results in only limited qualitative agreement with experiment.

The first question to settle is the stability of the nonmagnetic state of the system of holes and Mn atoms towards ferromagnetic state. This is determined by examining when the system is unstable when a small external static, constant magnetic field is applied. This instability is in general a function of hole density, concentration of Mn atoms, and the temperature. The temperature at which this instability occurs for given hole density and Mn-concentration is the Curie temperature. This question is addressed by an examination of the total static paramagnetic susceptibility of the system expressed in the form $\chi_{Total} = \chi_{hole} + 2\chi_{hole,Mn} + \chi_{Mn}$, where the individual response functions are given by $\chi_{hole} = \iint d\vec{r} d\vec{r}' \langle s_z(\vec{r}) s_z(\vec{r}') \rangle$, $\chi_{hole,Mn} = \sum_{l} \int dr \langle s_z(\vec{r}) S_z(l) \rangle$, and $\chi_{Mn} = \sum_{l,l'} \langle S_z(l) S_z(l') \rangle$. These response functions represent spin fluctuations and correlations of the hole spin and Mn spins, and Mn spin fluctuations. The incipient ferromagnetic instability of the system is signalled by a singularity of this total susceptiblity towards ferromagnetism and a criterion for this is obtained just as in the celebrated theory of ferromagnetism in the Stoner model. This will give the Curie temperature as a function of hole density and Mn concentration.

Next, anticipating the system to be ferromagnetic, we develop equations for the single particle spinor Green function of the holes based on the Hamiltonian given by eq.(1, 2, 3), which contains besides the mean-field contributions of the type mentioned above, a vertex contribution. Similarly, the equation determining the mean value of the local spin of the Mn is found to contain besides the Landau-Lifshitz type torque, contributions arising from the mean values of the other Mn-spins and the itinerant hole spin-polarization, another contribution arising from the same vertex that appeared in the equation for the hole-Green function. When the vertex function is calculated in leading order of the exchange coupling, J, it is found that these contributions incorporate fluctuations of localized Mn-spins as well as those of itinerant hole spin densities. At this stage of the theory, the fluctuations of the Mn-spins is also calculated to close the loop, the equations for which in turn contain all the other quantities. These then form the full self-consistent set of equations which, we believe, goes beyond the simple mean field theory mentioned above. The hole gas can be treated as a 3-dimensional bulk system or as a 2-dimensional system which is pertinent to the Field Effect Transitor arrangement (Rajagopal 1998).

The spinor hole Green function contains information about the single particle hole spectrum, the total number of holes as well as the hole-spin polarization. A nonzero value of the Mn-spin, $\langle S \rangle$, is central to the ferromagnetism of the whole semiconductor system. The equal time commutation relations among the components of the local Mn-spin operators, for example, $[S_x(l), S_y(l')] = iS_z(l)\delta(l,l')$ imply that the nonzero value of the expectation value of $\langle S \rangle$ requires a self-consistent determination of the Mn-spin fluctuations. The vertex functions, $\Gamma$, are measures of the interaction between Mn-spins and Mn-spin with hole-spin. The generalization of the mean field theory is thus seen to be essential in a realistic formulation of the ferromagnetism of doped semiconductors.

From this theory, we capture the RKKY-type interaction: the usual RKKY result is obtained when one examines the large-distance behavior of the interaction between two Mn-spins when the host system is magnetically unpolarized giving the traditional RKKY result and when it is magnetically polarized, recovering the result in (Kochalev 1964). In both of these cases, the Kohn singularity at twice the Fermi momentum of the hole gas residing in the hole-gas (spin) density fluctuation determines the RKKY form. These results are derived using 3-dimensional hole gas system. This corresponds to the high charge density situation mentioned earlier. In the low density limit, the result is very different and the double exchange result ensues. Our theory is able to interpolate between these two regimes, which is what is needed in the materials under investigation.

**B. Functional theory**

In order for feasible realistic calculations of magnetic semiconductor properties, the method of choice is the functional formulation of the above many-body theory. We will now outline briefly how this may be accomplished (Rajagopal 1998).

**(1) Mapping theorem for the stationary (time-independent) case** (when we examine the thermodynamic equilibrium): the density matrix and the free energy associated with the Hamiltonian given by eqs.(1, 2, 3) are functionals of the mean values of hole density, n(r), hole spin-density vector, $\underline{s}$(r), and the local spin vector, $\underline{S}$(l), of the impurity.

**(2) Mapping theorem for nonstationary (time-dependent) case** (when we examine the excited states of the system, such as the spin waves): the time dependent density matrix and the Action, A, are functionals of the time-dependent mean values of hole density, n(r,t), hole spin-density vector, $\underline{s}$(r,t), and the local spin vector, $\underline{S}$(l,t), of the impurity. From the action, we deduce the equations for the two ingredients of this system:

**(3) (a) Pauli-like equation for the hole-spinor**, which involves effective potential and effective magnetic field, both of which are functionals of the mean values of hole density, hole spin-density vector, and the local spin vector.

$$i\frac{\partial}{\partial t}\underline{\Psi} = \left\{-\frac{\overleftrightarrow{\nabla}^2}{2m^*} + V_c\left[n,\vec{s},\vec{S}\right] + \vec{\tau}\cdot\vec{W}_c\left[n,\vec{s},\vec{S}\right]\right\}\underline{\Psi} \tag{8}$$

These functionals are respectively the functional derivatives of the action functional with respect to hole density and hole spin density vector. The hole-density and the hole spin-density vector are expressed in terms of the solutions of the Pauli equation and thus a self-consistent set of equations.

**(3) (b) Landau-Lifshitz-like equation for the mean local spin vector**, where the effective magnetic field experienced by the local spin is another functional derivative with respect to the local spin vector of the action functional, and hence is also a functional of the three quantities mentioned in (3)(a).

$$\frac{\partial}{\partial t}\vec{S}(l) = \vec{H}_c\left[n,\vec{s},\vec{S}\right] \times \vec{S}(l) \qquad (9)$$

We note from these statements that the functional theory depends only on one functional, namely the action functional, from which all else follows. One therefore needs ways of setting up this functional. One of the ways of doing this is to look back at the results of the many-body theory in (A), and start with a suggestive form of the functional. Another method is a generalization of a method called the "PHI- DERIVABLE" method of the many-body theory which is based on using some known conservation laws in choosing some selected set of "diagrams" in setting up the functional.

The interactions between the local Mn-spins and the holes are captured in the action functional A from which the effective potential, $V_c$, experienced by the holes, which is a functional of n, $\varsigma, \langle S \rangle$, determine the excitation spectrum of the holes in a self-consistent manner. Similarly, the local magnetic spin of Mn, obeying the effective Landau-Lifshitz equation experiences an effective magnetic field, $H_C$, which is a functional of n, $\varsigma, \langle S \rangle$. In both of these equations, the mean local spin and the hole spin-density appear and are influenced by the interactions, exhibiting their dynamic decoherence on a scale of the strength of exchange interaction and the value of the local Mn-spin.

In both of these theoretical approaches, one has a set of self-consistent equations to solve. After that one computes physical quantities of interest, such as the Curie temperature, the spin-current density (Rajagopal and Mochena 2000), etc.

**C. Self-consistent numerical analysis based on A and B.**

Akai (1998) has published a first principles calculation using (KKR-CPA-LDA) density functional theory to examine the carrier induced ferromagnetism in diluted magnetic semiconductor (In,Mn)As. Our procedure will go well beyond what was done in this work, in that the local density functional (LDA) employed by Akai was based on electron gas results which is not suitable in the present context. Our program is to first examine the many-body theory, from which we will set up a functional appropriate to this system to serve as a proper basis for computation of the system properties. An important point to make here is that the Pauli-like equation mentioned above for the polarized hole-gas, leads automatically to two types of currents, one associated with particle density and another, with the vector-spin-density. The first obeys the usual continuity equation. The vector spin-density obeys a somewhat different form of continuity equation, containing now a spin-current density which is a tensor, along with a contribution from the Lorentz-force like contribution arising from the local spin vectors of the Mn.

## III. SUMMARY

The regions of instability of the nonmagnetic state towards the ferromagnetic state of a combined system of Mn-atoms in a semiconductor is first investigated. Then a self-consistent many-body theory of the ferromagnetic state is formulated. This goes beyond the current mean field theories in that it includes the fluctuations in the Mn-spins and the itinerant hole-gas of the semiconductor. It naturally leads to a development of a functional theory which is the method of choice for numerical implementation of many-body theory. Actual computation of system properties such as Curie temperature as a function of hole concentration and the Mn-concentration, spin-current transport coefficients, etc. of interest, based on the functional theory in both 3-D (bulk) and 2-D (pertinent to FET structure) are planned. This work will thus enable us to predict the suitable conditions under which one may obtain practical ferromagnetic semiconductor systems operating in the device configurations.

**Acknowledgements**

We thank Professor S. D. Mahanti of Michigan State University, for many valuable suggestions and for pointing out relevant references to the literature. Two of the authors (AKR and PJLC) are supported in part by the US Office of Naval Research.